\documentclass[twoside,11pt]{article}

\usepackage{obs_study_style}

\usepackage{amsmath}
\usepackage{booktabs}
\newcommand{\indep}{{\bot\negthickspace\negthickspace\bot}}
\newcommand{\ra}[1]{\renewcommand{\arraystretch}{#1}}
\usepackage{xcolor}

\heading{}{}{}{}{}{Pallavi Basu and Dylan S. Small}

\ShortHeadings{Constructing a More Closely Matched Control Group in a DID Analysis}{Basu and Small}
\firstpageno{1}

\begin{document}



\title{Constructing a More Closely Matched Control Group in a Difference-in-Differences Analysis: Its Effect on History Interacting with Group Bias}

\author{\name Pallavi Basu \email pallavi\_basu@isb.edu \\
       \addr Area of Operations Management\\
       Indian School of Business\\
       Hyderabad, Telangana 500111, India 
       \AND
       \name Dylan S. Small \email dsmall@wharton.upenn.edu \\
       \addr Department of Statistics\\
       The Wharton School, University of Pennsylvania \\
       Philadelphia, PA 19104, USA}

\maketitle


\begin{abstract}%
Difference--in--differences analysis with a control group that differs considerably from a treated group is vulnerable to bias from historical events that have different effects on the groups. Constructing a more closely matched control group by matching a subset of the overall control group to the treated group may result in less bias.  We study this phenomenon in simulation studies.  We study the effect of mountaintop removal mining (MRM) on mortality using a difference--in--differences analysis that makes use of the increase in MRM following the 1990 Clean Air Act Amendments. For a difference--in--differences analysis of the effect of MRM on mortality, we constructed a more closely matched control group and found a 95\% confidence interval that contains substantial adverse effects along with no effect and small beneficial effects.
\end{abstract}



\section{Introduction}

Mountaintop removal mining (MRM) mines coal within mountaintops by removing the mountaintop through clearcutting forests and explosives, and then surface mining the excavated coal.  The removed mountaintop material is deposited in valleys, burying streams.  There is evidence that MRM has negative environmental impacts, and there are concerns about human health (\citealt{palmer2010mountaintop,lindberg2011cumulative}).   MRM was introduced in the 1960s but was not widespread until the 1990s (\citealt*{szwilsld2000mountaintop,hendryx2016unintended}), increasing in part because of 1990 Clean Air Act (CAA) amendments that made coal reserves in the Central Appalachian region more financially attractive (\citealt{milici2000depletion,hendryx2016unintended}).  The acres directly impacted by MRM went from 71,046 and 77,100 in 1976 and 1985 respectively to 199,991 and 272,252 in 1995 and 2005 respectively based on satellite image analysis (\citealt{skytruth2009}).  Using the CAA amendments as a treatment that increased exposure to MRM, \citet{hendryx2016unintended} studied MRM's health impact using difference--in--differences analysis.  In Section 2, we examine the robustness of a related difference--in--differences analysis to violations of the parallel trend assumption by constructing a more closely matched control group.

Difference--in--differences analysis estimates a treatment effect by considering a treatment group receiving treatment in an after but not a before period and a control group not receiving treatment in either period and taking the difference between the difference of the treated group after and before outcomes and the difference of the control group after and before outcomes (\citealt*{meyer1995natural,cook2002experimental}; \citealt{grabich2015county};  \citealt*{wing2018designing}). Difference--in--differences unbiasedly estimates the causal effect of treatment if the treatment and control groups would have exhibited parallel trends in the counterfactual absence of treatment, and there is no interference among units nor multiple versions of treatment; see Figure 1. The parallel trends assumption could be violated if historical events in the after period affect the two groups differently {\textendash} history interacting with group bias (\citealt*{meyer1995natural,cook2002experimental}).  Concern about history interacting with group bias might be lessened if the control group is more similar to the treatment group.  We propose to use multivariate matching to construct a control group that is more closely matched to the treatment group on covariates that might affect outcomes. In Section 3, we study the impact of such matching in estimates of a treatment effect in a difference--in--differences analysis with the aid of numerical experiments.

\subsection{Background on MRM}

MRM is a form of surface mining conducted at the ridge of the Appalachian mountains. Increased demand for coal in the United States, sparked by the 1973 and 1979 petroleum crises, created incentives for a more economical form of coal mining than traditional underground mining methods which involves hundreds of workers, triggering the first widespread use of MRM. Its prevalence expanded further in the 1990s to retrieve relatively low-sulfur coal, a cleaner-burning form, which became desirable as a result of the 1990 CAA amendments that tightened emissions limits on high-sulfur coal processing (\citealt{burns2005}).  The development of bigger equipment that enabled MRM at scale in the later 1990s also drove increases in MRM (\citealt{bozorgebrahami2003}; \citealt{ramani2012}).


Mountaintop mined counties are among the poorest in the central and southern Appalachia; see Figure 4.  Historically, mining may have contributed to this poverty (\citealt{eller1982}), but by the 1970s, this poverty was well established (see Figure 4) and needs to be controlled for in any analysis of the effect of MRM on health. Following (\citealt{hendryx2016unintended}), we consider the effect of MRM on all-cause mortality and mortality from respiratory diseases.  We examine all-cause mortality because MRM could have effects on many aspects of health including cardiovascular disease and cancer through air and water pollution, and mental health through solastalgia and stress of health concerns (\citealt{hendryx2009}; \citealt{christian2011}; \citealt{canu2017}; \citealt{cordial2012}).  Mortality from respiratory diseases was selected because of prior evidence that MRM generate local air pollution and may promote poor health outcomes for these conditions (\citealt{kurth2014}; \citealt{hendryx2009}; \citealt{christian2011}).  


\subsection{Background on Historical Crises} 

Two historical events that occurred in the post-CAA amendment period that might have affected MRM county and control counties (counties in the Appalachian region without MRM) are the financial crisis of the late 2000s and the ongoing opioid crisis.  The financial crisis of the late 2000s led to the Great Recession and severely impacted almost all of North America, but disadvantaged areas tended to be hit harder (\citealt{thiede2016}; \citealt{burgard2015}). It is therefore highly plausible that the effect of the CAA amendment that led to increases in MRM by the late 2000s will be confounded with the impact of the Great Recession in more impoverished and disadvantaged counties.

The opioid crisis began in the late 1990s. The opioid crisis has tended to hit rural regions particularly hard (\citealt{compton2016}; \citealt{keyes2014}). We quantify the urbanization of the counties via the rural-urban continuum code of 1983, ranging from 0 to 9 with very high numbers indicating overly rural. We note that more than 80 percent of the MRM counties in the Appalachian region, has a rural-urban continuum code of 7 or higher when compared to about 46 percentage of MRM absent counties in the area.

We consider a difference-in-differences study with 1979--1989 as pre-period and 1990--2016 as post periods. The usual difference-in-differences method assumes that historical events in the post period, such as the opioid crisis and the Great Recession, did not differentially affect the exposed group (MRM counties) and control group.  We will demonstrate in simulation studies that when there is concern about a differential effect, multivariate matching on covariates that might be associated with such a differential effect, e.g., socioeconomic characteristics, can reduce bias. First, we review previous research on available methodologies for using matching in difference-in-differences studies.  

\subsection{Review of Available Methodologies}

In this section, we compare and contrast several existing methodologies for conducting difference-in-differences studies with multivariate matching. One concern in using matching to construct a control group in a difference-in-differences study is regression to the mean (RTM). In the careful analyses by \citet{daw2018}, they have noted that when the pre-period response level is matched on and is correlated with treatment assignment, then matched studies are biased. This is because in the post-period outcomes of the units both from the treatment and the matched control groups regress to their group means, creating a superficial difference between the groups even when the null hypothesis of no treatment effect holds.  In our situation, we do not match with pre-period response levels but match with covariates that we believe correlates with the differential impact of the post-period historical event on the outcome and assume that these covariates are not related to random fluctuations in the pre-period outcome.  \citet{daw2018} discuss that their study does not consider the type of matching we consider and that the type of matching we consider might be beneficial:~``In this study, we do not consider the possibly beneficial application of matching on covariates that differ between the groups and are correlated with future outcome trends (i.e., matching on confounders in the difference-in-difference sense). Instead, we consider matching on covariates that are correlated with levels of the outcome (or are the outcome level itself), which is the source of the regression to the mean bias we discuss here." To illustrate this point further, we conducted a version of the second numerical setting with no matching, full matching, and matching with pre-period responses. The results are reported in Table 8 and discussed further in Appendix B.

There is work on using propensity scores in matching in difference-in-differences studies (\citealt{stuart2014}) but it does not fully address the setting that motivates our work. Our version of matching involving rank–based Mahalanobis distance with calipers on the propensity score has not been studied in the DID framework.  In Stuart et al.~(2014), the weights are obtained by estimating the propensity scores of being in any of the four groups given the covariates. The four groups in their work are defined by both time and intervention status. Because of the formulation, the authors' have noted that their work is more appropriate when the composition of each group may change over time. Their work requires an exogeneity condition on the covariates, which is not essential in our work. Furthermore, the motivation of our work is different, being that potential historical events interacting with groups, not mentioned in their work.

Prognostic scores are an alternative to propensity scores that summarize covariates' association with potential responses (\citealt{hansen2008}).  A drawback of prognostic score based methods for our setting is that the model for the prognostic score is based only on the control group. We think this would be relevant for us, as seen even from our simulation settings when the outcome model for the treatment and the control groups are different. Additionally, we have not found a thoughtful study of the prognostic score-based methods in the DID literature to make a careful comparison.


The synthetic control methodology constructs a weighted combination of control units whose weighted pre-period observations and covariates are close to a treated unit (\citealt{abadie2010}).  Generally, researchers have used the synthetic control method when there are a small number of treatment units and multiple observations in the pre-period. The three original examples from Abadie et al. [2003, 2010, 2015] and the R package Synth (\citealt{abadie2011}), deal with one treatment unit (Basque Country, California, and West Germany respectively) with about twenty or more pre-treatment outcomes. In theory, synthetic control may be applied to multiple treatment units. According to \citet{abadie2020}, ``estimation with several treated units may carry some practical complications." We refer the interested readers to Section 8 of an excellent review by \citet{abadie2020}. We have applied matching with many treated units, e.g., the MRM application to which we apply matching has 32 treated counties. And matching seeks to find control units which are individually close to treated units rather than construct a weighted combination of control units.  Another difference between the synthetic control methodology and the matching approach we study in this paper is that the synthetic control methodology uses past outcomes in addition to covariates to construct the synthetic control.  The use of past outcomes can make the synthetic control method vulnerable to bias from regression to the mean (\citealt{illenberger2020}; see also the discussion in \citet{kaul2015}).

\citet{bernal2018} suggests the controlled interrupted time series (ITS) or CITS to overcome the disadvantage that a basic ITS has, namely, ``can not exclude confounding due to co-interventions or other events occurring around the time of the intervention." The authors suggest that the idea is to discover a control series that limits the impact of historical bias. Our approaches are ``complementary" to this in the sense that we aim to find a control paired unit so that the bias due to historical event is alleviated. As discussed above, synthetic control can be thought of a way of finding an imaginary comparative unit, and the authors have further noted in a letter to the editor (\citealt{bernal2019}) that indeed that the use of synthetic controls is complementary and ``not an alternative approach to CITS." \citet{linden2018} discusses the use of synthetic control in ITS. Likewise, one can use our version of matching in conjunction with ITS.

\section{Effect of Mountaintop Removal Mining on Mortality}

\subsection{Methods}

We consider the effect of MRM on age--adjusted all--cause and respiratory disease-specific mortality rates per 100,000 persons obtained from the Centers for Disease Control and Prevention (CDC). We use 1979--1989 as the before period and after periods divided into 1990--1998 and 1999--2016.  The before period was chosen to be a period in which mortality rates were relatively stable and the after period was divided into two parts because MRM may have lagged effects (\citealt{hendryx2016unintended}).  The time periods chosen were also selected based on data availability considerations; see Section 2.3. We consider the Appalachian counties as defined by the \citet{appalachianregionalcommission}.  The treated group is the 32 counties where there was MRM between 1994--2006 using \citet{esch2011chronic}'s classification based on satellite imagery and the control group is the 396 counties without MRM between 1994--2006.

The MRM counties differ considerably from the controls in the before period with lower income, higher poverty and lower education; see Table \ref{results.table}.  Historical events in the after period that might affect health include the Great Recession and opioid
 crisis. The opioid crisis and the economic recession might interact with the factors that differ between the MRM and control groups in affecting health.  For example, the opioid crisis may be differentially distributed between the treated and control counties threatening the counterfactual assumption of parallel trends in the post period.  If one can explain the differential distribution of the opioid crisis across the two groups by measured factors, then one can hope to mitigate this problem.  If possible, it would be useful to test how predictive our matching variables are of the distribution of the post-period factor among the treatment groups to assess whether matching has helped to reduce differences in the post-period factor.  Unfortunately we do not have available data on the distribution of the opioid crisis in different counties.  Another way in which the opioid crisis might interact with the factors that differ between the MRM and control groups in affecting health is that the opioid crisis may have the same distribution between the treated and control counties, but counties with poorer populations may have fewer resources to cushion the blows from it and be harder hit by it.  If one can match on the poverty level of a county, then one can hope to mitigate potential violations of the counterfactual assumption of parallel trends in the post-period from poorer populations having fewer resources to cushion blows from the opioid crisis.

We constructed a more closely matched control group by pair matching MRM to control counties using rank--based Mahalanobis matching with calipers on the propensity score (\citealt{rosenbaum2010dos}), using the \texttt{optmatch} package in R (\citealt{hansen2006optmatch}).  The fourth column of Table \ref{results.table} shows the matched controls are much closer to the MRM counties on the covariates.  We tried matching each MRM county to two control counties, but the balance was not acceptable (see Table \ref{supplementary.balance.table}).

\subsection{Counterfactual Assumptions for Identifiability}

We index the $N$ Appalachian counties by $i$ where $i = 1, \ldots, N$, and let the treatment, that is, the prevalence of MRM, be indicated by $W_i$ taking values in $\{0, 1\}$. Let $p=0$ denote the before period and $p=1$ denote the after period. Let $Y_{ip}(1)$ denote the potential outcome under treatment, and $Y_{ip}(0)$ denote the counterfactual outcome in the absence of treatment. Let $X_{ip}$ denote a vector of $d$-dimensional covariates. In our setup, we assume that all the counties considered have observations available in both periods. We assume that there is no interference between the counties and that the stable unit treatment value assumption (SUTVA) holds.

We are interested in estimating the population average treatment effect for the treated (PATT) in the after period, in notations, $E \{ Y_{i1}^{W_i = 1}(1) - Y_{i1}^{W_i = 1}(0) \} $, the empirical analogue for which is $\frac{1}{|\{i:~W_i = 1\}|} \sum_{i:W_i = 1}  \{ Y_{i1}(1) - Y_{i1}(0) \} $. The difficulty in this case is that the quantity $Y_{i1}^{W_i = 1}(0)$ is unobserved. Since in the pre-period, the treatment is not applied, if there is a model based framework, we can potentially apply a modeling approach to find an estimator $\widehat{Y_{i1}}^{W_i = 1}(0)$. However, in the absence of a good available model due to our failure to anticipate all future events, and to enable model-free estimation, the information on the control counties are utilized.  When the parallel trend assumption (i.e., the trends of the treated and control groups would have been parallel in the absence of treatment) holds that $E_1(Y_{i1}^{W_i = 1}(0) - Y_{i0}^{W_i = 1}) = E_1(Y_{i1}^{W_i = 0}(0) - Y_{i0}^{W_i = 0})$ upholds, this enables $\widehat{Y_{i1}}^{W_i = 1}(0) := Y_{i0}^{W_i = 1} + \{ Y_{\cdot 1}^{W_{\cdot} = 0}(0) - Y_{\cdot 0}^{W_{\cdot} = 0} \}$, where $E_1(\cdot)$ indicates that the expectation is taken with respect to the post period. Note that $Y_{\cdot 1}^{W_{\cdot} = 0}(0)$ is observed.

Now suppose that there is a historical event $H$ that concurs with the treatment under study between the pre and post periods. We observe the outcomes $Y_{i1}^{W_i = 1, H = 1}(1)$ and $Y_{i1}^{W_i = 0, H = 1}(0)$ in the after-period. In presence of the unavoidable historical event, we still aim to estimate the PATT that is, $E \{ Y_{i1}^{W_i = 1, H = 1}(1) - Y_{i1}^{W_i = 1, H = 1}(0) \} $. However, even if the the counterfactual parallel trends assumption would have held in the absence of the historical event, we can no longer generally assert that $E_1(Y_{i1}^{W_i = 1, H = 1}(0) - Y_{i0}^{W_i = 1}) = E_1(Y_{i1}^{W_i = 0, H = 1}(0) - Y_{i0}^{W_i = 0})$ holds if there is a differential impact of the (future) historical event. To remove such a differential impact of historical events, ideally we would like to form a matched pair $(i, i')$ such that $W_i = 1, W_{i'} = 0$ and $E_1(Y_{i1}^{W_i = 1, H = 1}(0) - Y_{i0}^{W_i = 1}) = E_1(Y_{i'1}^{W_{i'} = 0, H = 1}(0) - Y_{i'0}^{W_{i'} = 0})$ holds. In terms of the familiar \textit{unconfounded} assumptions we require that 
\begin{equation} 
E_1(Y_{i1}^{H = 1}(0) - Y_{i0}|W_i = 1, X_{i0}) = E_1(Y_{i1}^{H = 1}(0) - Y_{i0}|X_{i0}) \label{diff.in.diff.unconfounded}
\end{equation} 
holds. Different matching strategies are discussed in the numerical experiments section and is implemented in the MRM study. Our modified difference-in-differences estimator for the PATT, aided by matching, is formulated as $\frac{1}{|\{i:~W_i = 1\}|} \sum_{(i, i'):~W_i = 1, W_{i'} = 0}  \{ (Y_{i1}^{H = 1}(1) - Y_{i0}) - (Y_{i'1}^{H = 1}(0) - Y_{i'0})\}$. For matching based on assumption (\ref{diff.in.diff.unconfounded}) to be able to eliminate bias,  we need the assumption of overlap or probabilistic assignment of treatment, where every unit in the pre-period has a positive chance of being assigned as a treatment unit or a control unit in the post-period.

We discuss the comparison of our matched DID assumption to the assumption of \textit{unconfounded assignment} (Imbens and Rubin 2015), which is the assumption that treatment is independent of potential outcomes given the measured covariates. A stronger version of the usual DID methodology assumes that the post-period counterfactual outcome levels conditional on the pre-period outcomes are independent of the treatment assignment. The stronger version of our assumption is that the same holds but in the absence of any conjunctive historical events, in notations $(Y_{i1}^{H = 0}(1) - Y_{i0}, Y_{i1}^{H = 0}(0) - Y_{i0})  \indep W_i$, in the notation of \citet{dawid1979}. In the presence of historical event we modify the unconfounded assignment assumption to $(Y_{i1}^{H = 1}(1) - Y_{i0}, Y_{i1}^{H = 1}(0) - Y_{i0})  \indep W_i \vert X_{i0}$, where recall that $X_{i0}$ is the vector of $d$-dimensional covariates at time period zero.

\subsection{Implementation}

{\it{Outcome}}.  Our outcome is age--adjusted all--cause mortality rates and age-adjusted mortality rates by diseases of the respiratory system obtained from the CDC.  Mortality rates are per 100,000 persons and are age--adjusted to the 2000 US population.  The CDC WONDER website's compressed mortality files enable mortality rates to be calculated for 1968--1978, 1979--1998, and 1999--2016, or any subset of the years within each of these time periods.  We chose to make 1979--1989 the before period because mortality rates were relatively stable in this period after declining from 1968--1978 (\citealt{hendryx2016unintended}).  We divided the after period into the two periods 1990--1998 and 1999--2016 because MRM might affect mortality with a lag (\citealt{hendryx2016unintended}).

{\it{Treated and Control Groups}}.  The counties we considered are the Appalachian counties as defined by the \citet{appalachianregionalcommission}.
We used \citet{esch2011chronic}'s classification of which counties have MRM which was based on satellite imagery.

{\it{Covariates}}.  \citet{lengerich2004appalachia} identified education, isolation, access to health care, health beliefs, and kinship as factors affecting mortality in Appalachia (\citealt{lengerich2004appalachia,lengerich2006images}).  We sought to control for some of these factors by matching on the following county-level covariates: income, poverty rate, demographics (percent white and percent female) and education (percent with no high school degree, percent with bachelors degree), poverty rate and rural-urban continuum code. Median income, poverty rate, percent white, percent female, percent with no high school degree and percent with bachelors degree were taken from the 1990 US Census. Our pre-period is from 1979–1989, and the two post-periods are 1990–1998 and  1999–2016. Therefore we have chosen the 1990 census data for our covariates. We checked the change in distribution in poverty rates, one of the socioeconomic covariate, displayed in Figure 4. Particularly in the control counties, there does not seem many visual differences from 1980 to 1990 to 2000 to 2010. Furthermore, the treatment may directly or indirectly affect some covariates, such as employment or education. We want to balance the treatment and the control counties in these covariates at or around the time when the treatment commenced. The rural--urban continuum code was taken from the 1983 county rural--urban continuum codes from the USDA Economic Research Service where 9 is most rural and 0 is most urban (\citealt{ruralurban1983}).

{\it{Analysis}}.  We used the R package \texttt{optmatch} (\citealt{hansen2006optmatch}) to pair match MRM counties to control counties on the covariates using rank--based Mahalanobis matching with a caliper on the logit of the propensity score of  0.2 $\times$ the standard deviation of the logit of the propensity score, following the matching method suggested in Chapter 8 of \citet{rosenbaum2010dos}. We tried matching each MRM county to two control counties, but the balance was not acceptable, e.g., the standardized difference for poverty rate was 0.49; see Table \ref{supplementary.balance.table}.

We made inferences about the treatment effect from the matched pairs data as follows.  Let $y_i$ for the $i$th matched pair be the difference between the difference of the MRM county's after and before outcome (age adjusted mortality rate per 100,000) and the difference of the control county's after and before outcome.  Each $y_i$, $i=1,\ldots ,32$ is an estimate of the treatment effect for the $i$th matched pair.  We model the $y_i$ as independent and identically distributed normal random variables with mean equal to the treatment effect and find a confidence interval for the treatment effect by inverting the $t$--test. For the unmatched data, we make inference for the treatment effect by regressing the outcome for each county in each time period on a time period dummy, a fixed effect for the county, and a dummy variable for whether MRM was being used for the county in the time period (the coefficient on this variable estimates the treatment effect) (\citealt{imbens2009recent}).


\subsection{Results}

The bottom panel of Table \ref{results.table} shows the difference--in--differences estimates for the effect of the increase in MRM starting in 1990 on age--adjusted mortality rate (per 100,000) using all controls and matched controls.  For the later after period (1999--2016), using all controls, we estimate the increase in MRM increased the mortality rate by 94 (95\% CI: 62, 126) while using closely matched controls, we estimate it increased the death rate by 28 (95\% CI: -8, 64).  The closely matched control analysis suggests the all controls analysis may have overstated the effect of MRM but also suggests further study of MRM’s effect on health is warranted, with a 95\% confidence interval that contains substantial adverse effects even though it also contains no effect and small beneficial effects.

Table \ref{results.table.2} shows the difference--in--differences estimates for the effect of the increase in MRM starting in 1990 on age-adjusted mortality rate (per 100,000) from diseases of the respiratory system using all controls and matched controls.  For the later after-period (1999--2016), using all controls, we estimate the increase in MRM increased the respiratory cause-specific mortality rate by 9.70 (95\% CI: 2.66, 16.76) while using closely matched controls, we estimate it increased the respiratory cause-specific death rate by -2.38 (95\% CI: -12.78, 8.01).

\section{Simulation Study}
We consider two simulation setups aimed to reflect what may happen in a simplistic real--data scenario. In both the situations the observed effect or response and the incidence of the treatment are related to the covariates of the subjects, sometimes fully observed while sometimes only partially noted. We describe the data generating process and explain the numerical results in the rest of this section.

\subsection{Simulation 1}
In the first setting we assume that the vector of covariates of the control population is generated from $Z_C \sim N(\mu_C, \Sigma_C)$ with $\mu_C := (1, \dots, 1)^T$ and $\Sigma_C$, the variance--covariance matrix is such that the variance is set to $1$, and the covariance is assumed to each take a value 0.2. This may be viewed as a weak--positive correlation between the covariates. The  covariates of the treatment population are then assumed to come from $Z_{Tr} \sim N(\alpha_{Tr}\mu_C, \Sigma_{Tr})$, with our choice of $\alpha_{Tr} := 1.25$ and $\Sigma_{Tr} := \Sigma_C$. In words, the covariates of the treatment population appear to come from a right--shifted Gaussian curve, with substantial overlap with that of the covariates of the control group. Further, the response for the control and the treatment subjects are assumed to be structured as below for a two time--period framework,
\[
Y_{C,it} = \beta_{0,C} + \beta_{1,C} * T + \varepsilon_{it},
\]
\[
Y_{Tr,jt} = \beta_{0,Tr} + \beta_{1,Tr} * T + \varepsilon_{jt}.
\]
Where $T := 0$ indicates the before treatment or event period, and $T := 1$ for the after period and $\varepsilon_{it}$ and $\varepsilon_{jt}$ denote independent standard Gaussian variables. In our specific simulation setting we choose $\beta_{0,C} = \beta_{0,Tr} = 0$. Finally the slope variables relate to the covariates by a linear transformation, specifically, $\beta_{1,C} := \beta^T Z_C$ and $\beta_{1,Tr} := \beta^T Z_{Tr} + \Delta$, with the entries of $\beta \sim \text{Unif}[2,3]$ and $\Delta := 2$. The goal for us is to recover the treatment effect $\Delta$ from the observed responses when a) all the covariates are known, and b) only half of all the covariates are observed and are available for matching. We fix the treatment and the control sample sizes to 32 and 320 respectively. The number of covariates is varied within $\{2, 4, 8\}$. For the covariate size of eight, a stylistic plot of the setup is shown in Figure 2.

Is a $\Delta =2$ a reasonable effect size to recover? We perform a back-of-the-envelope calculation for the setting with the covariate size of two. Our post-period outcome is defined as $Y = \beta_0 + \beta_1 * \{T = 1\} + \varepsilon$ where $\beta_0 \equiv 0$. The slope coefficient $\beta_1 = \beta^T Z_C$ for the control group and for the treatment group is $\beta_1 = \beta^T Z_{Tr} + \Delta$ where $Z_C$ and $Z_{Tr}$ are the covariates of the respective groups. Both $Z_C$ and $Z_{Tr}$ are generated from a multivariate normal with variance-covariance matrix $\Sigma$ with diagonal entries all 1 and off-diagonal entries as 0.2. The entries of the vector are independently and identically distributed as $\beta \sim Unif[2,3]$ and $\Delta \equiv 2$ is a constant. We note that for each of the treatment or the control groups for the post-period $Var(Y) =   Var(\beta^T  Z) + 1$. Note that by the law of total variance $Var(\beta^T  Z) \geq E(Var(\beta^T  Z)|\beta) = E(\beta^T Var(Z) \beta)$. Suppose we work with the minimum covariate size in the performed simulation that is covariate size of 2. Denote $\beta = [U_1, U_2]^T$ where $U_1, U_2 \sim Unif[2,3]$. Therefore $ Var(\beta^T  Z) \geq E(U_1^2 + U_2^2 + 2 \cdot 0.2 \cdot U_1 \cdot U_2) \approx 15.17 $ and $Var(Y) \geq 16.17$. Therefore the standard deviation of the post-period responses in each groups are greater than 4, and in the convention of effect sizes see, for example, \cite{sawilowsky2009}, our study choice would be of medium to small effect size. This involves thus a challenging setting even in the best-case scenario. Given the approximations and conservativeness of the calculation, we have tried for $\Delta = 6$, and the performance is almost similar. 

All results are reported by implementing 1000 replications of the above--explained setting. In this setting, we hope that the bias will be much less when matching with all the available covariates, and less when matched with only half of the available covariates. In addition to reporting the mean and the median estimates, we also report the `coverage proportion,' the ratio of the 1000 replications in which the constructed 95\% confidence intervals contained the true $\Delta$.

In Table 4 we note that with an increase in the matching of the covariates, the mean and the median estimates are closer to the truth. The coverage proportion also greatly improves with matching. Even with partial matching, the coverage proportion is above 0.85, a considerable improvement over being not matched at all. Note that the mean and the median estimates are both biased from the truth when there is no matching and when matched with only half of the covariates. Furthermore, the general difficulty of the problem is magnified by the dimension of the covariates, given that the sample sizes are fixed. In Table 5 we fix the number of covariates to 4 and vary the equicorrelation of covariates within \{0.1, 0.05, 0\}. The additional observation to note here is that the more uncorrelated the variables are, the less well matching with partial covariates performs.

\subsection{Simulation 2}
In this setting we assume that both the control and the treatment covariates marginally arise from $Z \sim N(\mu, \Sigma)$, or from the same covariate population with $\mu := (1, \dots, 1)^T$ and $\Sigma$ is the same as that defined in Experiment 1. Here, however, the conditional chance of being selected in the treatment is related to the covariates by $P(Tr = 1|Z) := 1 - \Phi(\beta^T Z)$. In this setting, the entries of $\beta \sim \text{Unif}[.2,.3]$, so that only about 10\% of the population is therefore allocated to treatment. Now we model the potential impact of a historical event, by considering that a historical event has an impact with a chance $P(\text{Historical Event Impacts} = 1|Z) := 1 - \Phi(\beta^T Z)$. Here again, the parameters are chosen such that about 10\% of the population is on an average impacted by the historical event. Note, though the two chances greatly depend on the covariates, in any realization the treatment population and the population being impacted by the historical event may be the same or may be entirely different. We begin with 400 subjects in total, which includes both control and treatment. The size of the available covariates is fixed at 8.

The responses are also generated differently from what we did in Experiment 1. In the before--period, the responses for both the control and the treatment are from a standard Gaussian. In the after--period, in the absence of an impact of the historical event and in the absence of the treatment, the response is again from a standard Gaussian. If the historical event has an impact, but still in the absence of a treatment, the response is shifted from the standard Gaussian by an amount of $\delta$. If the historical event does not have an impact but the treatment is applied, the response deviates from a standard Gaussian by $\Delta$. In the presence of both, that is, both that the treatment is applied and the historical event has an impact, the response drifts from the standard Gaussian by an enlarged--or could be diminished--amount of $\delta + \Delta$. Our goal is to obtain an accurate estimate of the treatment effect $\Delta$ which teases apart the effect of the treatment from the impact of the historical event and the noisy responses that is, distorted with Gaussian random variables. Note though that we do not have direct information on whether the historical event had an impact or not, or its effect size, other than what we can retrieve from the observed responses and the given covariates. Here again, we study the setting for when a) all the covariates are available for matching, and b) only half of all the covariates are available to match. The specific choice of $\Delta = -2$ and that for $\delta$ is varied in $\{-2, -4, -6\}$ to reflect the size of impact of the historical event. Table 6 reports a summary of the results of 1000 replications. Here again, we can visualize a stylistic figure similar to figures 1 and 2, where the usual DID estimator will be on the order of $\Delta + \delta$, whereas the estimand of interest is $\Delta$. A multivariate covariate-based matched estimator will approximately equate the estimand.

In this setting, we hope that the bias will be lessened by matching in finite samples. Asymptotically we expect that the bias is removed with full matching. To see that, we provide some heuristics. Matching the i'th treatment subject to the j'th control implies matching $p_{i} := P(H_i = 1|Tr_i = 1, Z_i)$ and $p_{j} := P(H_j = 1|Tr_j = 0, Z_j)$, where the event $\{H_i = 1\}$ indicates that there is an impact on the i'th subject by the historical event. The amount of unit bias is given by $(1/n) \Sigma_{i} \{ H_i - p_i \} - (1/n) \Sigma_{j} \{ H_j - p_j \} + (1/n) \Sigma_{(i, j) = (1, 1)}^{(n, n)} \{ p_i - p_j \}$, where $n$ denotes the number of treatment samples matched to one unique control each, $H_i$ and $H_j$ indicate the realized (unknown to the analyst) values of the impact or nonimpact of the historical event, and the enumeration $(i, j) = (1, 1)$ to $(n, n)$ stands for $(i, j) = (1, 1), (2, 2), \dots, (n, n)$. We assume that as a result of matching, we are able to obtain $(1/n) \Sigma_{(i, j) = (1, 1)}^{(n, n)} \{ p_i - p_j \} = o(1)$, where the asymptotics is in the sense that a large number of controls are increasingly available to match to. The independence and mean zero characteristics of $\{H_i - p_i\}$ and $\{H_j - p_j\}$ then ensure that both $(1/n) \Sigma_{i} \{ H_i - p_i \} \rightarrow 0$ and $(1/n) \Sigma_{i} \{ H_i - p_i \} \rightarrow 0$ hold almost surely. Combining these three terms eliminates the bias asymptotically. This is empirically demonstrated in Table 7 where the effective sample size is increased to about twelvefold to that of the treatment sample size in Table 6 in the setting described above.

Table 6 provides some more insights in addition to largely following the patterns of Tables 4 and 5. The coverage proportion for the partial matching hovers around 0.80, that is, away from the desired value of 0.95, while the coverage proportions for the unmatched analysis is much worse, ranging from 0.06 to 0.49. Here, note the bias is not completely removed, in finite samples by matching but it is reduced. Better matching strategies, perhaps using more structure of the generating process may reduce the bias with a potential increase in the variance. The performance in the larger sample size of Table 7 with full matching is reasonable with coverage proportions above .9 and nearly unbiased estimates.

\section{Conclusion and Limitations}
Difference--in--differences analysis with a control group that differs considerably from a treated group is vulnerable to bias from historical events that have different effects on the groups.  Constructing a more closely matched control group by matching a subset of the overall control group to the treated group may result in less bias.  For a difference--in--differences analysis of the effect of mountaintop removal mining on mortality, we constructed a more closely matched control group and found a 95\% confidence interval that contains substantial adverse effects along with no effect and small beneficial effects. 

Our matching methodology is very general and can be applied whenever there is a suspicion that other events are impacting both the control and the treatment groups. The other requirement is to have covariates available for matching that allows a level playing field for the comparisons of the desired outcomes. There are several instances of such studies in the literature. One, an investigation of the impact of a vacant lot greening program in Philadelphia on health and safety outcomes (\citealt{branas2011}). Another is a study on estimating the effect of the repeal of comprehensive background check laws on firearm homicide and suicide rates in Indiana and Tennessee (\citealt{kagawa2018}). And a third study of the impact of the revised women, infants, and children (WIC) food package on maternal nutrition during pregnancy and postpartum (\citealt{hamad2019}).

Our study has limitations.  We were unable to obtain pre-treatment period measures of health, such as smoking and obesity. We also did not match on other factors that might affect mortality in Appalachia, such as isolation, access to health care, employment, health beliefs, and kinship (\citealt{lengerich2004appalachia}).  We considered a county to be a control unit if the county itself did not have MRM, but counties adjacent to MRM counties might suffer spillover effects from environmental pollutants. Such spillover effects would bias our estimate of MRM's effects on mortality downwards.

\bigskip

\textbf{Acknowledgements.} The authors thank the editor and the two anonymous reviewers for many comments that considerably improved the work. Pallavi Basu is grateful to Prof.~Yoav Benjamini for intriguing her interests in the ideas in epidemiology and observational studies and to the Wharton School for providing excellent facilities and friendly work conditions during her visit.

\newpage

\appendix

\section{Test for Parallel Trend in the Pre-Period}

In the standard DID model (we refer to an excellent review by Imbens and Wooldridge, 2009) the outcome for observation $i$ in the absence of the intervention is formulated as
\[
Y_i(0) = \alpha + \beta \cdot T_i + \gamma \cdot G_i + \varepsilon_i,
\]
where $T_i = \{0, 1\}$ refers to the time indicator, and $G_i = \{0, 1\}$ refers to the group indicator where the treatment group is coded as 1. In the presence of the intervention, researchers model the outcome as
\[
Y_i(1) = Y_i(0) + \tau_{did}.
\]
One estimates the parameter $\tau_{did}$ by regressing the observed outcome on the time, group, and their interaction indicator, namely,
\[
Y_i = \alpha + \beta_1 \cdot T_i + \gamma_1 \cdot G_i + \tau_{did} \cdot T_i \cdot G_i + \varepsilon_i'.
\]

We implemented a parallel trend test through visual exploration and a test of significance of $\tau_{did}$ in the pre-period. The p-value of the coefficient is 0.381, indicating statistically not significant. In Figure 3, we plot the mean outcome of total mortality for the four periods. The four periods are 1968-1978, 1979-1989, 1990-1998, and 1999-2016. Qualitatively, we do not see evidence that parallel trends is violated in the pre-periods that is between 1968-1978 and 1979-1989, and the larger than the usual nominal level of p-value supports that.

\section{Matching on Outcomes and Regression-to-the-Mean}

In Table 8, we follow a part of the setting of numerical experiment 2. However, we demonstrate the performance of our proposed (full) multivariate covariate-based matching when compared to the matching by pre-treatment outcomes. This is an important topic as matching with the pre-treatment outcomes may display a regression-to-the-mean (RTM) phenomenon. We make one change to make the point. We simulate $\beta_{cov} \sim Unif[.2, .3]$, and for every response at all time points the outcomes are defined as $Y_{it}' := Y_{it} + \beta_{cov}^TZ_{it}$, where $Y_{it}$ are the outcomes as defined in Section 3.2. Since we do not change the covariates with time we may as well write $Z_{i0} = Z_{i1} =: Z_{i}$.

We observe in Table 8 that matching on the pre-period outcomes gives us worse or more biased estimators than not matching at all. The reasoning is the following. By design, our treatment units, on average, have very low values of the covariates. The pre-period responses involve matching the response values at the initial time, which also by design will be lower for the treatment units. The matched control units, therefore, to be well-matched with outcomes, have to either have very low covariates or very low realizations in the Gaussian noise, that is, the random component, or a combination of the two aspects. This is critical to understanding why large bias occurs.  Since there is an aspect of the matching that involved the larger-than-usual random component of the outcome in the pre-period of the control units, in the post-period, most of them will regress to its expected value. The superficial difference created by the regression contributes to the bias and is, therefore, systematically more biased than not matching at all. We do not recommend matching units based on the outcomes.

\begin{center}
\includegraphics[scale=0.644]{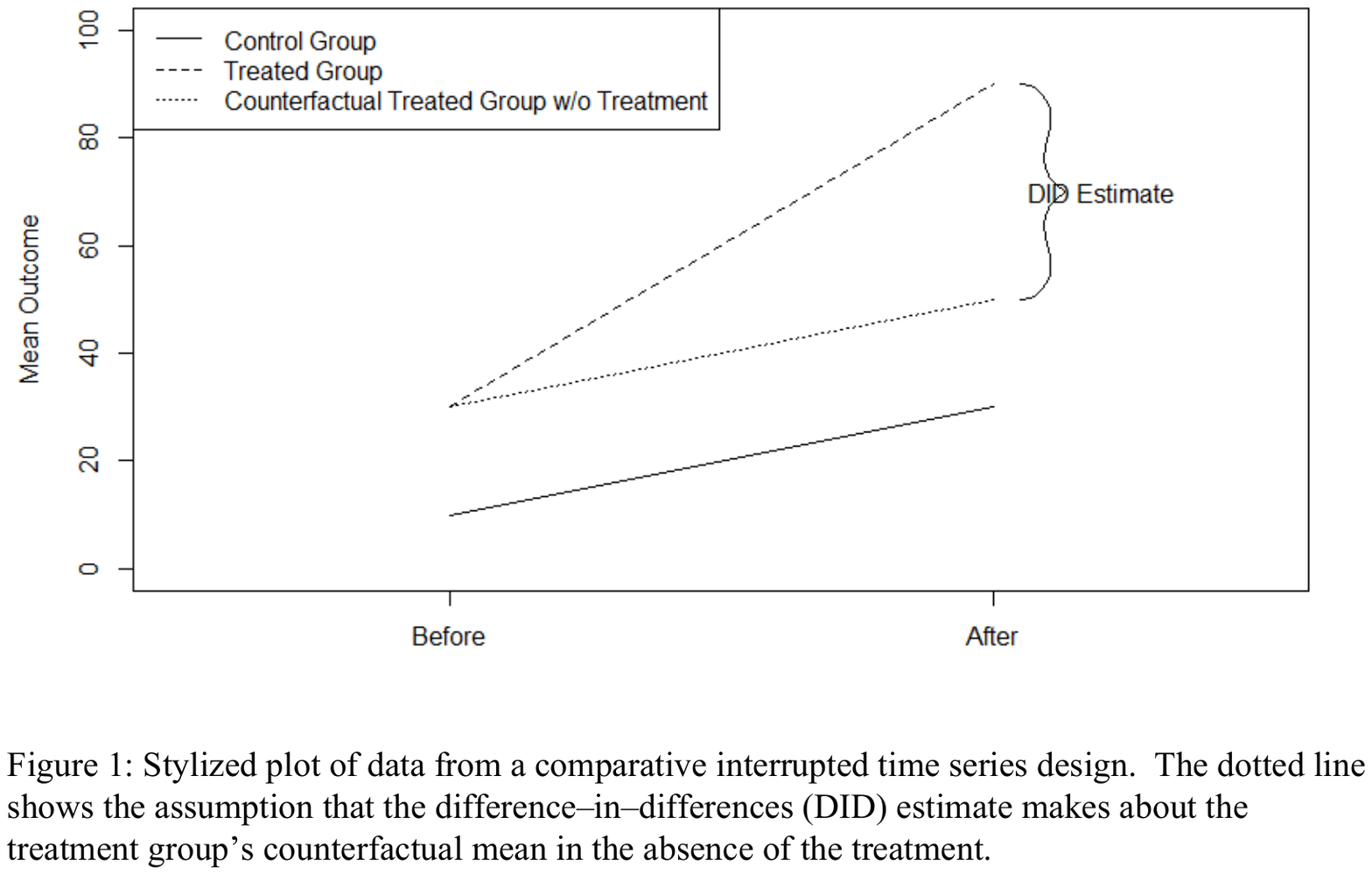}
\end{center}

\begin{table}[h!]
\caption{Covariate balance between mountaintop removal mining (MRM) counties and all control counties (all counties in the Appalachian region that do not have MRM) and pair matched control counties as well as age--adjusted mortality rates (per 100,000; adjusted to 2000 U.S. population) and difference--in--differences estimates of effect of increase in MRM starting in 1990 on age adjusted mortality per 100,000 persons using all controls and pair matched controls.  The covariates matched on are the rural--urban continuum code in 1983 (the last year before 1990 in which it was measured), which ranges from 0 to 9 with 9 being the most rural \citep{ruralurban1983}, and the following county--level variables measured at the end of the before period from the 1990 US Census:~median income, poverty rate, percent white, percent female, percent with no high school degree, percent with high school degree, and percent with bachelors degree.  The before matching (Bf.~Mt.) standardized difference (Stand.~Diff) is the differences in before matching group means divided by the before matching within group standard deviation (specifically the square root of the average of the before matching within group variances).  The after matching (Af.~Mt.) standardized difference (Stand.~Diff) is the differences in after matching group means divided by the before matching within group standard deviation (\citealt{rosenbaum1985constructing}).}
\begin{center}
\begin{tabular}{|c|c|c|c|c|c|}
\hline \multicolumn{6}{|c|}{Covariates} \\
\hline & MRM Counties & All Controls & Matched Controls & \multicolumn{2}{|c|}{Stand. Diff.} \\
& & & & Bf. Mt. & Af. Mt. \\ \hline
Income (Median) & 8,111 & 10,291 & 7,913 & -1.38 & 0.12 \\
Poverty Rate (\% ) & 31.1 & 18.1 & 30.6 & 1.78 & 0.06 \\
White (\% ) & 97.8 & 93.0 & 98.0 & 0.59 & -0.03 \\
Female (\% ) & 51.4 & 51.4 & 51.5 & -0.03 & -0.13 \\
No High School Degree (\% ) & 50.4 & 37.9 & 51.2 & 1.45 & -0.10 \\
High School Degree (\%) & 41.8 & 51.3 & 41.7 & -1.41 & 0.02 \\
Bachelors Degree (\% ) & 7.8 & 10.8 & 7.1 & -0.74 & 0.17 \\
Rural--Urban Continuum Code & 7.2 & 5.8 & 7.5 & 0.69 & -0.17 \\
\hline \multicolumn{6}{|c|}{Outcomes} \\ \hline
Age Adjusted Mortality Rate & & & & &  \\
\hspace{5mm} 1979--1989 & 1,141 & 1,022 & 1,078 & & \\
\hspace{5mm} 1990--1998 & 1,115 & 976 & 1,056 & & \\
\hspace{5mm} 1999--2016 & 1,134 & 921 & 1,042 & & \\ \hline
\hline \multicolumn{6}{|c|}{Difference--in--Differences Estimates (95\% CI) of Treatment Effect} \\ \hline
Treatment Period & & All Controls & Matched Controls & & \\ \hline
1990--1998 & & 20 (-1, 41) & -4 (-32, 23) & & \\
1999--2016 & & 94 (62, 126) & 28 (-8, 64) & & \\ \hline
\end{tabular}
\end{center}
\label{results.table}
\end{table}

\begin{table}
\caption{Covariate balance between mountaintop removal mining (MRM) counties and matched control counties when two control counties were matched to each MRM county. The standardized difference is the differences in after matching group means divided by the before matching within group standard deviation (specifically the square root of the average of the before matching within group variances).}
\begin{center}
\begin{tabular}{|c|c|c|c|}
\hline Covariate & MRM Counties & Matched Controls & Standardized Difference\\ \hline
Income (Median) & 8,111 & 8,526 & -0.26 \\
Poverty Rate (\% ) & 31.1 & 27.5 & 0.49 \\
White (\% ) & 97.8 & 97.8 & 0.00 \\
Female (\% ) & 51.4 & 51.7 & -0.32 \\
No High School Degree (\% ) & 50.4 & 47.5 & 0.34 \\
High School Degree (\% ) & 41.8 & 44.5 & -0.40 \\
Bachelors Degree (\% ) & 7.8 & 8.0 & -0.06 \\
Rural--Urban Continuum Code & 6.7 & 6.4 & 0.22 \\ \hline
\end{tabular}
\end{center}
\label{supplementary.balance.table}
\end{table}

\begin{table}[h!]
\caption{Age-adjusted mortality rates (per 100,000; adjusted to 2000 U.S. population) by diseases of the respiratory system and difference--in--differences estimates of the effect of the increase in MRM starting in 1990 on age-adjusted mortality by diseases of the respiratory system per 100,000 persons using all controls and pair-matched controls.}
\begin{center}
\begin{tabular}{|c|c|c|c|}
\hline & MRM Counties & All Controls & Matched Controls \\
\hline \multicolumn{4}{|c|}{Age Adjusted Mortality Rate by diseases of the respiratory system} \\ \hline
Time Period & & &   \\ \hline
\hspace{5mm} 1979--1989 & 114.5 & 80.2 & 94.8  \\
\hspace{5mm} 1990--1998 & 132.4 & 97.0 & 116.0  \\
\hspace{5mm} 1999--2016 & 143.4 & 99.5 & 126.1  \\ \hline
\hline \multicolumn{4}{|c|}{Difference--in--Differences Estimates (95\% CI) of Treatment Effect} \\ \hline
Treatment Period & & All Controls & Matched Controls  \\ \hline
1990--1998 & & 1.16 (-3.98, 6.31) & -3.17 (-11.37 5.02)  \\
1999--2016 & & 9.70 (2.66, 16.76) & -2.38 (-12.78, 8.01)  \\ \hline
\end{tabular}
\end{center}
\label{results.table.2}
\end{table}

\clearpage

\begin{figure}
\begin{center}
\includegraphics[scale=0.8]{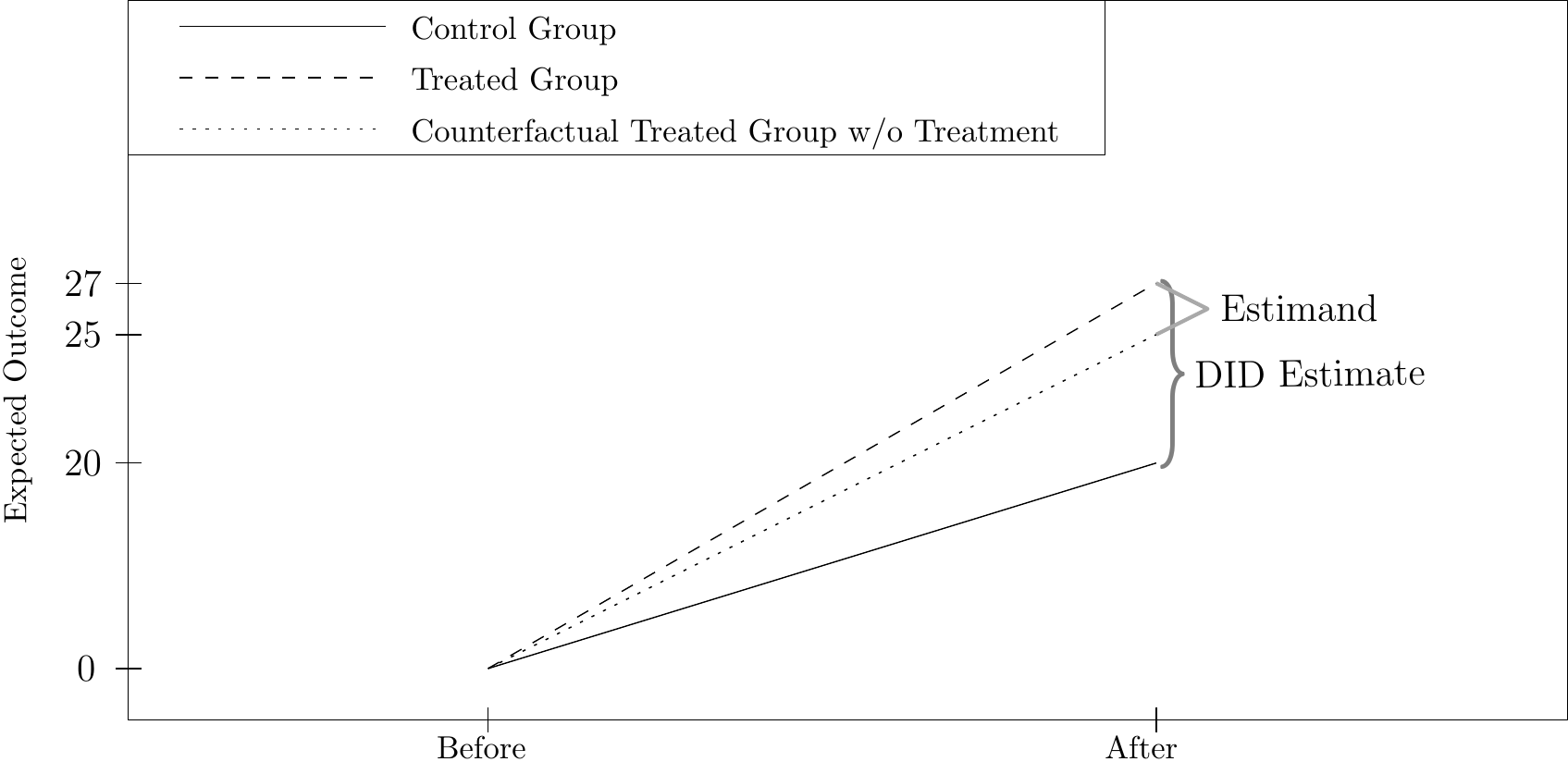}
\end{center}
\caption{A stylized plot of setting 1 of our numerical experiments for a covariate size of eight. The vanilla DID estimate ignores the impact of historical events interacting with groups and is biased. The dotted line shows the assumption that the matched estimator makes about the treatment group's counterfactual mean in the absence of the treatment. The matched estimator is less biased.}
\end{figure}

\begin{table*}\centering
\caption{Results for the first numerical experiment in setting 1 for the true difference--in--differences effect size of value $\Delta = 2$. We used the R package \textit{optmatch} to pair match treatment to control subjects using the rank--based Mahalanobis distance of the covariates with a caliper on the propensity score of 0.2 $\times$ the standard deviation of the logit of the propensity score. The acronyms SD, MAD, and CI stand for standard deviation, median absolute deviation, and confidence interval respectively. The entries are rounded to two decimal places.}
\ra{1.3}
\begin{tabular}{@{}lccc@{}}\toprule
& \multicolumn{3}{c}{Proportion of Covariates Matched}\\
\cmidrule{2-4}
& None & Half & Full\\ \midrule
\textbf{Number of Covariates = 2}\\
Mean (SD) & 3.24 (0.77) & 2.53 (0.73) & 2.01 (0.37)\\
Median (MAD) & 3.25 (0.73) & 2.51 (0.75) & 2.02 (0.38)\\
Coverage proportion (Length of CI) & 0.63 (3.00) & 0.90 (2.89) & 0.95 (1.51)\\
\textbf{Number of Covariates = 4}\\
Mean (SD) & 4.48 (1.21) & 2.83 (1.03) & 2.03 (0.55)\\
Median (MAD) & 4.47 (1.21) & 2.87 (1.03) & 2.03 (0.52)\\
Coverage proportion (Length of CI) & 0.46 (4.72) & 0.87 (4.06) & 0.95 (2.20)\\
\textbf{Number of Covariates = 8}\\
Mean (SD) & 7.01 (2.04) & 3.29 (1.56) & 1.89 (1.11)\\
Median (MAD) & 7.02 (2.08) & 3.36 (1.52) & 1.93 (1.07)\\
Coverage proportion (Length of CI) & 0.32 (8.04) & 0.87 (6.23) & 0.97 (4.90)\\
\bottomrule
\end{tabular}
\end{table*}

\begin{table*}\centering
\caption{Results for the \textbf{second} numerical experiment in setting 1 for the true difference--in--differences effect size of value $\Delta = 2$. We used the R package \textit{optmatch} to pair match treatment to control subjects using the rank--based Mahalanobis distance of the covariates with a caliper on the propensity score of 0.2 $\times$ the standard deviation of the logit of the propensity score. The acronyms SD, MAD, and CI stand for standard deviation, median absolute deviation, and confidence interval respectively. The entries are rounded to two decimal places. Values in bold in the entries indicate the notable observations in addition to those noted in Table 4.}
\ra{1.5}
\begin{tabular}{@{}lccc@{}}\toprule
& \multicolumn{3}{c}{Proportion of Covariates Matched}\\
\cmidrule{2-4}
& None & Half & Full\\ \midrule
\textbf{Equicorrelation = 0.1}\\
Mean (SD) & 4.53 (1.11) & \textbf{3.03} (0.95) & 2.01 (0.50)\\
Median (MAD) & 4.49 (1.14) & 3.05 (0.95) & 2.03 (0.48)\\
Coverage proportion (Length of CI) & 0.37 (4.29) & 0.83 (4.00) & 0.96 (2.06)\\
\textbf{Equicorrelation = 0.05}\\
Mean (SD) & 4.52 (1.06) & \textbf{3.10} (0.98) & 2.04 (0.49)\\
Median (MAD) & 4.51 (1.06) & 3.12 (0.96) & 2.03 (0.45)\\
Coverage proportion (Length of CI) & 0.32 (4.04) & 0.82 (3.97) & 0.95 (1.97)\\
\textbf{Equicorrelation = 0}\\
Mean (SD) & 4.51 (0.98) & \textbf{3.24} (0.99) & 2.06 (0.46)\\
Median (MAD) & 4.49 (0.99) & 3.25 (0.94) & 2.06 (0.45)\\
Coverage proportion (Length of CI) & 0.25 (3.78) & 0.77 (3.87) & 0.95 (1.91)\\
\bottomrule
\end{tabular}
\end{table*}

\begin{table*}\centering
\caption{Finite sample performances of the numerical experiment in setting 2 for the true difference--in--differences effect size of value $\Delta = -2$. The total number of subjects is 400 with about 10\% being treatment subjects. We used the R package \textit{optmatch} to pair match treatment to control subjects by the propensity score metric using the information on the covariates. The acronyms SD, MAD, and CI stand for standard deviation, median absolute deviation, and confidence interval respectively. The entries are rounded to two decimal places.}
\ra{1.3}
\begin{tabular}{@{}lccc@{}}\toprule
& \multicolumn{3}{c}{Proportion of Covariates Matched}\\
\cmidrule{2-4}
& None & Half & Full\\ \midrule
\textbf{Impact of historical event $\delta$ = -2}\\
Mean (SD) & -2.54 (0.31) & -2.27 (0.39) & -2.14 (0.40)\\
Median (MAD) & -2.54 (0.31) & -2.30 (0.39) & -2.15 (0.41)\\
Coverage proportion (Length of CI) & 0.49 (1.05) & 0.89 (1.57) & 0.92 (1.58)\\
\textbf{Impact of historical event $\delta$ = -4}\\
Mean (SD) & -3.09 (0.41) & -2.54 (0.52) & -2.25 (0.56)\\
Median (MAD) & -3.09 (0.39) & -2.54 (0.51) & -2.26 (0.54)\\
Coverage proportion (Length of CI) & 0.12 (1.24) & 0.84 (2.08) & 0.91 (2.11)\\
\textbf{Impact of historical event $\delta$ = -6}\\
Mean (SD) & -3.61 (0.57) & -2.82 (0.68) & -2.37 (0.72)\\
Median (MAD) & -3.60 (0.56) & -2.82 (0.67) & -2.37 (0.73)\\
Coverage proportion (Length of CI) & 0.06 (1.50) & 0.79 (2.77) & 0.90 (2.80)\\
\bottomrule
\end{tabular}
\end{table*}

\begin{table*}\centering
\caption{Large sample performances of the numerical experiment in setting 2 for the true difference--in--differences effect size of value $\Delta = -2$. The total number of subjects is 2000 with about 24\% being treatment subjects. We used the R package \textit{optmatch} to pair match treatment to control subjects by the propensity score metric using the information on the covariates. The acronyms SD, MAD, and CI stand for standard deviation, median absolute deviation, and confidence interval respectively. The entries are rounded to two decimal places. Values in bold in the entries indicate the notable observations in addition to those noted in Table 6.} 
\ra{1.5}
\begin{tabular}{@{}lccc@{}}\toprule
& \multicolumn{3}{c}{Proportion of Covariates Matched}\\
\cmidrule{2-4}
& None & Half & Full\\ \midrule
\textbf{Impact of historical event $\delta$ = -2}\\
Mean (SD) & -2.19 (0.08) & -2.07 (0.11) & \textbf{-2.02} (0.10)\\
Median (MAD) & -2.19 (0.09) & -2.07 (0.11) & -2.02 (0.10)\\
Coverage proportion (Length of CI) & 0.38 (0.32) & 0.88 (0.40) & 0.94 (0.40)\\
\textbf{Impact of historical event $\delta$ = -4}\\
Mean (SD) & -2.37 (0.11) & -2.14 (0.13) & \textbf{-2.04} (0.13)\\
Median (MAD) & -2.37 (0.11) & -2.14 (0.12) & -2.04 (0.12)\\
Coverage proportion (Length of CI) & 0.05 (0.38) & 0.80 (0.49) & 0.93 (0.49)\\
\textbf{Impact of historical event $\delta$ = -6}\\
Mean (SD) & -2.56 (0.15) & -2.22 (0.16) & \textbf{-2.08} (0.16)\\
Median (MAD) & -2.55 (0.15) & -2.22 (0.16) & -2.08 (0.17)\\
Coverage proportion (Length of CI) & 0.01 (0.46) & 0.71 (0.61) & 0.91 (0.61)\\
\bottomrule
\end{tabular}
\end{table*}


\begin{figure}
\begin{center}
\includegraphics[scale=0.8]{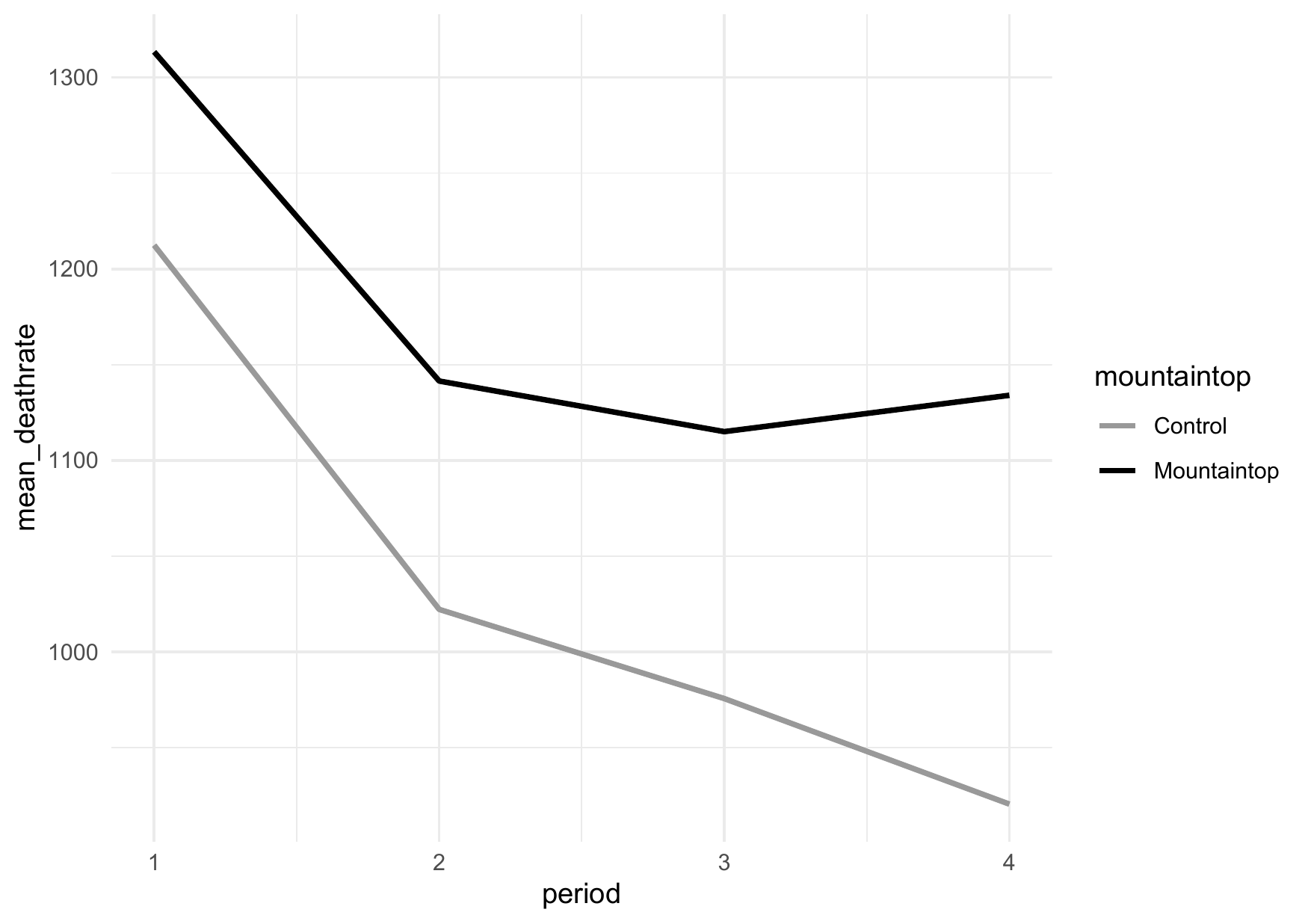}
\end{center}
\caption{Visual exploration of the parallel trend assumption in the control and MRM counties in the Appalachian region.}
\end{figure}

\begin{table*}\centering
\caption{A demonstration of finite sample performances of various matching strategies in the numerical experiment setting 2 for the true difference--in--differences effect size of value $\Delta = -2$. The total number of subjects is 400, with about 10\% being treatment subjects. We used the R package \textit{optmatch} to pair match treatment to control subjects by the propensity score metric using the information on the covariates. The acronyms SD, MAD, and CI, stand for standard deviation, median absolute deviation, and confidence interval, respectively. We round the entries to two decimal places.
}
\ra{1.3}
\begin{tabular}{@{}lccc@{}}\toprule
& \multicolumn{3}{c}{Matching Strategies}\\
\cmidrule{2-4}
& None & Proposed & To Pre-Period Outcome\\ \midrule
\textbf{Impact of historical event $\delta$ = -2}\\
Mean (SD) & -2.54 (0.30) & -2.12 (0.40) & -3.15 (0.38)\\
Median (MAD) & -2.55 (0.30) & -2.11 (0.41) & -3.14 (0.38)\\
Coverage proportion (Length of CI) & 0.47 (1.05) & 0.94 (1.61) & 0.16 (1.53)\\
\textbf{Impact of historical event $\delta$ = -4}\\
Mean (SD) & -3.09 (0.40) & -2.25 (0.54) & -3.51 (0.52)\\
Median (MAD) & -3.07 (0.41) & -2.25 (0.54) & -3.49 (0.53)\\
Coverage proportion (Length of CI) & 0.11 (1.23) & 0.92 (2.14) & 0.22 (2.14)\\
\textbf{Impact of historical event $\delta$ = -6}\\
Mean (SD) & -3.60 (0.56) & -2.35 (0.72) & -3.87 (0.71)\\
Median (MAD) & -3.59 (0.55) & -2.33 (0.70) & -3.85 (0.73)\\
Coverage proportion (Length of CI) & 0.06 (1.47) & 0.91 (2.75) & 0.25 (2.76)\\
\bottomrule
\end{tabular}
\end{table*}


\clearpage

\begin{figure}
\begin{center}
\includegraphics[scale=0.8]{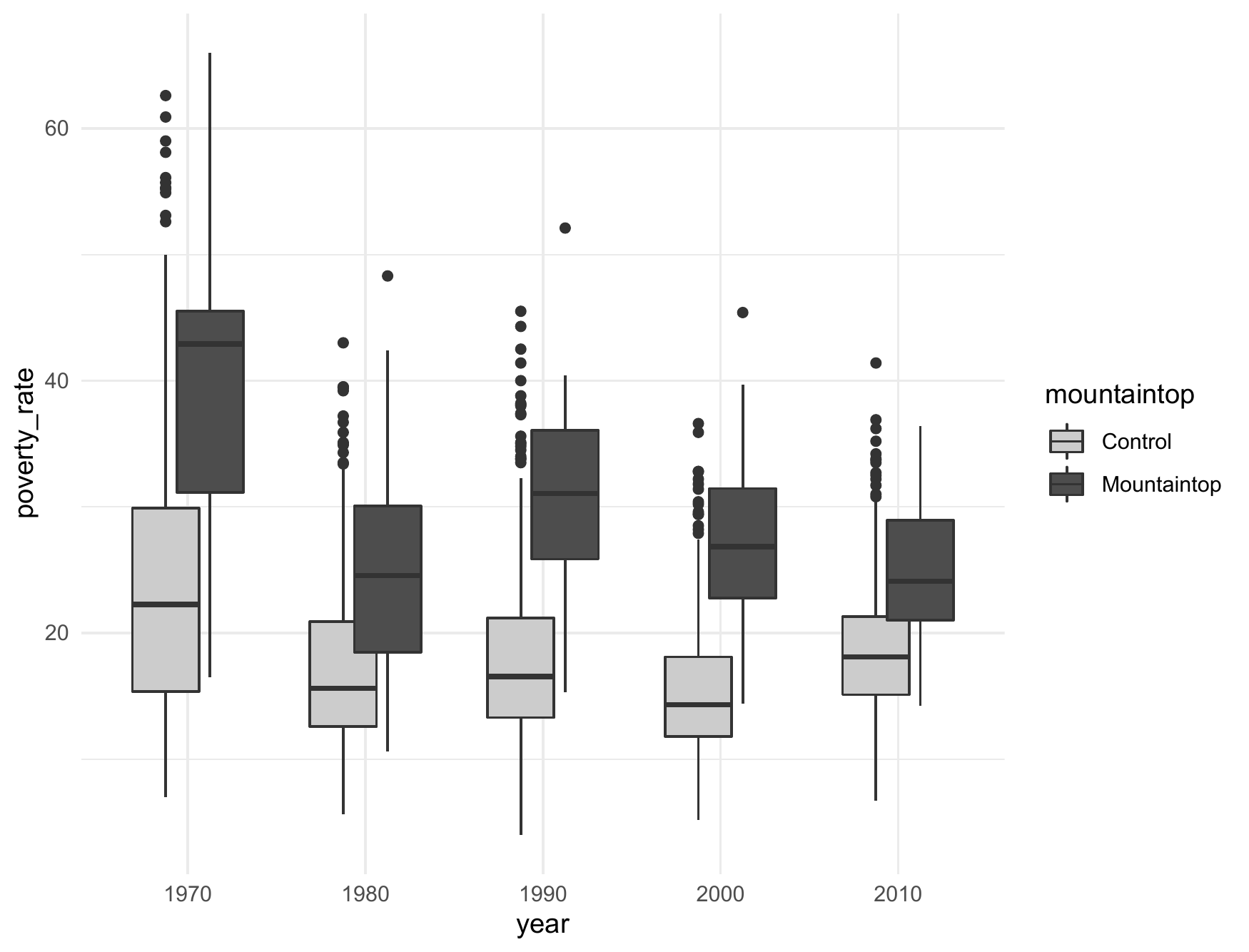}
\end{center}
\caption{Time evolution of poverty rates in the control and MRM counties in the Appalachian region.}
\end{figure}

\clearpage

\bibliographystyle{abbrvnat}

\bibliography{Manuscript}

\end{document}